\documentclass[preprint]{elsarticle}
\usepackage{amssymb}
\journal{Annals of Physics}
\begin{document}
\begin{frontmatter}
\title{Dust-acoustic rogue waves in an opposite polarity dusty plasma featuring non-extensive statistics}
\author{D. M. S. Zaman$^*$, A. Mannan, and A. A. Mamun}
\address{Department of Physics, Jahangirnagar University,
Savar, Dhaka-1342, Bangladesh\\ E-mail: saadzamanshaon@gmail.com$^*$}
\begin{abstract}
  Modulational instability (MI) of dust acoustic waves (DAWs), which propagates in an opposite polarity dusty plasma
  system, containing inertial warm negatively and positively charged dust particles as well as non-extensive q-distributed
  electrons and ions, has been theoretically investigated. The nonlinear  Schr\"{o}dinger (NLS)
  equation is derived by employing the reductive perturbation method. The NLS equation leads to the MI of DAWs as well as to the formation of DAW
  rogue waves (DARWs), which are formed due to the effects of nonlinearity in the propagation of
  DAWs. Both stable and unstable regions are revealed from the analysis of the NLS equation.
  It is observed that the basic features of the DAWs (viz. stability of the wave profile, MI growth rate, amplitude, and width of DARWs) are significantly
  modified by the various plasma parameters such as non-extensive
  parameter, electron number density, and electron temperature. The existence of the non-extensive electron/ion distribution
  creates an influence on the MI of the waves. It is
  observed that non-extensive distributed ions have more effect
  on the MI of the DAWs than electrons.
\end{abstract}
\end{frontmatter}
\section{Introduction}
A low-temperature plasma including dust particles with sizes ranging from $1$ to $10^3$ $\mu$m, is usually
referred to as dusty or complex plasma. In recent years, dusty plasmas (DPs) have received enormous attention
due to their crucial role in supporting electrostatic density perturbations and potential structures that are
observed in planetary rings \cite{Shukla2002,Mamun20002,Chowdhury2017a}, cometary tails \cite{Shukla2002,Mamun20002}, early universe
 \cite{Shukla2002,El-Taibany2013}, supernova \cite{Shukla2002,El-Taibany2013}, interstellar cosmic matter \cite{Shukla2002,El-Taibany2013}, dense
molecular clouds \cite{Shukla2002,El-Taibany2013}, Martian atmosphere \cite{Shukla2002,El-Taibany2013}, earth’s mesosphere \cite{Shukla2002,Mamun20002}, and laboratory experiments
 \cite{Shukla2002,El-Taibany2013,Zhukhovitskii2015}, etc .
In addition, DPs have recently enabled to achieve the importance in industries due to the contaminants formed
 during the plasma processing (deposition, etching, etc.) of thin films. The opposite polarity dust grains co-exist in both
space \cite{El-Taibany2008,El-Labany2013} and laboratory plasmas \cite{Ali1998,Zhao2002}, which introduce a new
dusty plasma model known as opposite polarity dusty plasma. Opposite polarity (positive and negative) dusts are
its main ingredients of opposite polarity dusty plasma \cite{Mamun20002,Mamun2015}. Definitely, depending on the
plasma environments, the significant or insignificant number of free electrons and ions are always found, which
are left over after their absorption by negative and positive dust grains \cite{Mamun2008,Mamun2016}. Dust grains
are normally regarded as negatively charged because they collect electrons from the background
plasma \cite{Shukla2002,El-Taibany2013}. However, the existence of positively charged dust particles has also been
noticed in different regions of space (viz. cometary tails \cite{El-Taibany2013}, Jupiter’s magnetosphere \cite{Mendis1991}, Earth’s polar mesosphere \cite{Chow1993}, Martian
atmosphere \cite{Horanyi1993}, and some plasma experiments \cite{Shukla2006,Angelo2004}, etc.).
Three principal mechanisms, i.e. photoemission under ultraviolet (UV) photons, thermionic emission induced by radiative
 heating, and secondary emission of electrons from the surface of the dust grain, are observed by which a dust grain acquires positive charge \cite{Mamun2002}.

As the Maxwellian distribution is inappreciable to describe systems in a non-equilibrium state with long range interactions,
particles of space and laboratory plasmas don't follow Maxwellian distribution in all time. In space and astrophysical
environments, the particles are supposed to follow the non-Maxwellian distribution, such as, non-extensive
$q$-distribution, when the plasma particles move very fast compared to their thermal velocities \cite{Ismael2017,Rafat2015,Shalini2015}.
Non-extensive generalization of the Boltzmann-Gibbs-Shannon entropy was first presented by Renyi \cite{Renyi1955} and
subsequently proposed by Tsallis \cite{Tsallis1988}, which has achieved enormous attention during last few decades.
The non-extensive distribution is suitable for the statistical description of long-range interaction systems, such as,
plasma and dusty plasmas. Non-extensive plasmas have been a fascinating research topic due to its relevance in
cosmological and astrophysical scenarios (viz. stellar polytropes, hadronic matter and quark-gluon plasma, dark-matter
halos, Earth’s bow-shock, magnetospheres of Jupiter and Saturn, etc. \cite{Plastino1993,Gervino2012,Feron2008,Asbridge1968,Krimigis1983})
as well as laboratory applications like nano-materials, micro-devices, and micro-structures \cite{Vladimirov2004}, etc.

Moreover, an unexpected, rare, and mysterious collective behavior which is known as rogue waves (also called freak waves, etc)
has been observed in many plasma systems. Rogue waves are short lived but high
energy event, which appear suddenly, and increase up to a very high amplitude (two, three, or even more times the
height of the surrounding waves), and finally disappear without any trace. It was first observed in ocean \cite{Akhmediev2009},
and also found later in optical systems \cite{Moslem2011,MSE2011}, fiber optics \cite{Moslem2011,MSE2011}, Bose-Einstein condensates \cite{Solli2007},
superfluid helium \cite{Kibler2010}, optical cavities \cite{Bludov2009},
atmospheric physics, and plasma physics \cite{Stenflo2010,Chowdhury2017} and even
in biology and stock market crashes \cite{Turing1952,Yan2010}. However, a possible mechanism to understand the rogue
waves is the rational solution of nonlinear Schr\"{o}dinger (NLS) equation.

Recently, Bains \textit{et al.} \cite{Bains2013} studied nonlinear self-modulation of low-frequency electrostatic dust acoustic
waves (DAWs) propagating in a dusty plasma system. El-Taibany \cite{El-Taibany2013} investigated nonlinear DAWs in a four
component inhomogeneous dusty plasma system. To the best of our knowledge, no theoretical investigation has been made to study the rogue waves in a
four-component dusty plasma system in the presence of non-extensive electrons and ions by deriving the NLS equation. However,
in our present work, we will examine the modulational instability (MI) of the DAWs propagating
in such kind of opposite polarity dusty plasma system (containing inertial warm negatively and positively charged dust
particles), as well as non-extensive $q$-distributed electrons and ions, which abundantly observed in astrophysical environments.

The paper is organized in the following fashion: The model equations and derivation of NLS equation are
presented in Sec. \ref{sec:eqn}. The stability of DAWs and rogue waves are presented in
Sec. \ref{sec:Stability}. The discussion is provided in Sec. \ref{sec:Discussion}. The Appendix: Expressions
of the coefficients is included in Sec. \ref{sec:Appendix}.
\section{The Model Equations and Derivation of the NLS equation}
\label{sec:eqn}
\begin{figure}[t!]
\centering
\includegraphics[width=70mm]{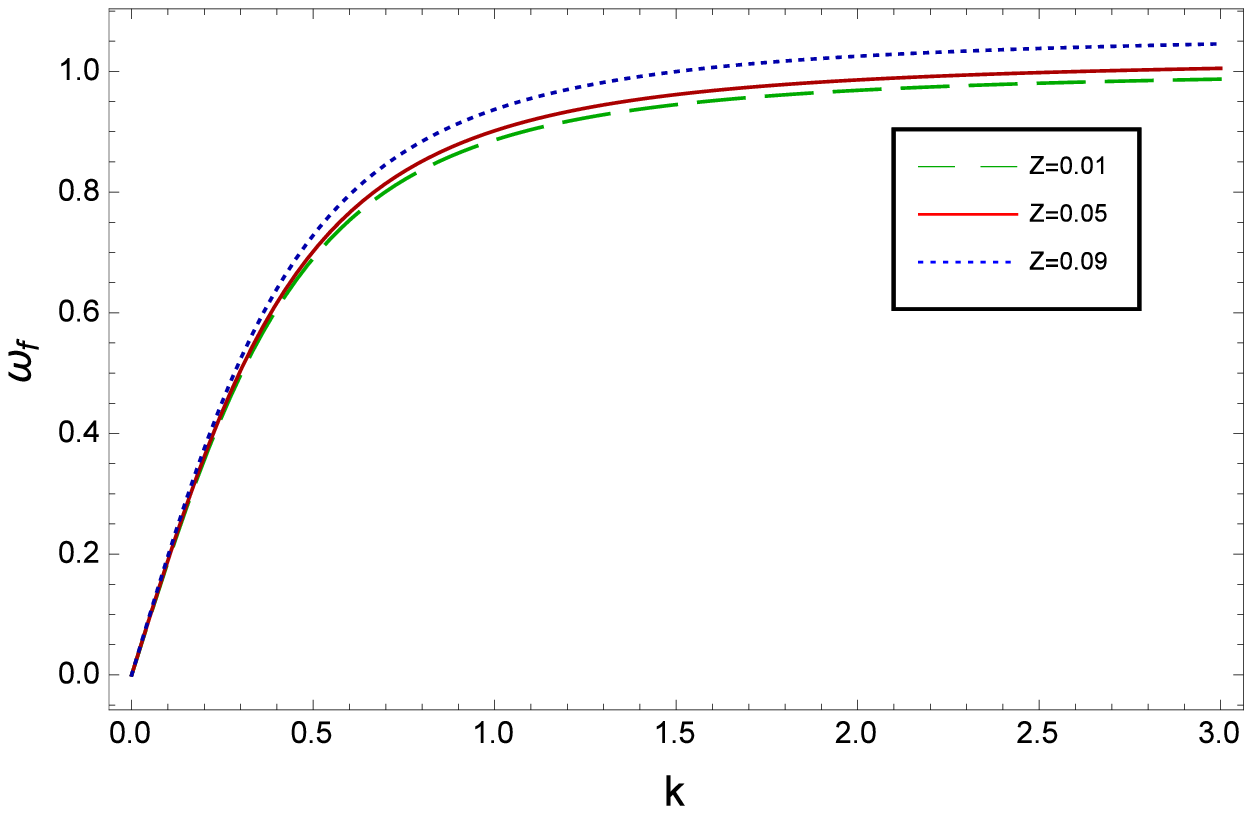}
\caption{The variation of $\omega_f$ with $k$ for different values $Z$;
 along with $\mu=150$, $R=0.1$, $\mu_i=0.4$, $\delta=0.3$,  $q=1.5$, $\sigma_1=0.0001$, and $\sigma_2=0.001$.}
\label{1Fig:F1}
\vspace{0.8cm}
\includegraphics[width=70mm]{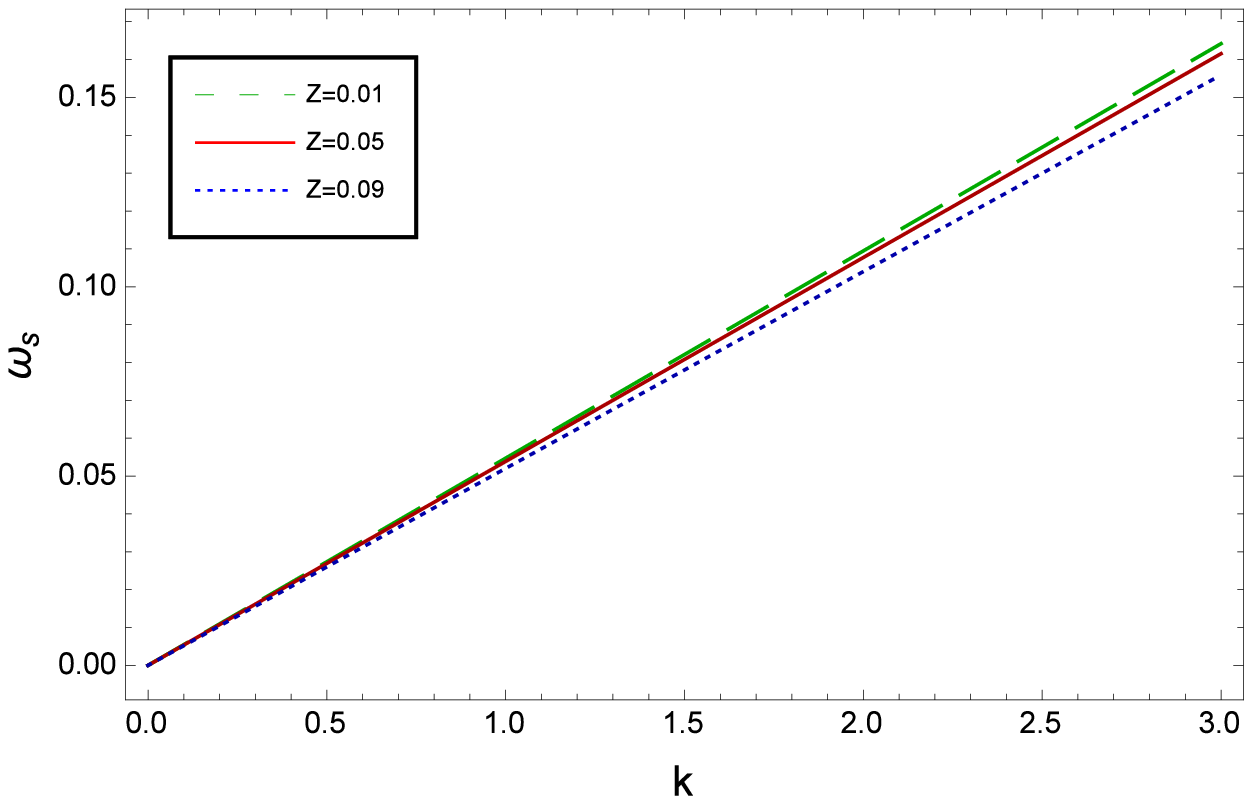}
\caption{The variation of $\omega_s$ with $k$ for different values $Z$;
 along with $\mu=150$, $R=0.1$, $\mu_i=0.4$, $\delta=0.3$,  $q=1.5$, $\sigma_1=0.0001$, and $\sigma_2=0.001$.}
 \label{1Fig:F2}
\end{figure}

We consider a collisionless, fully ionized four-component unmagnetized plasma system consisting of
inertial warm negatively charged dust particles and positively
charged dust particles as well as non-extensive $q$-distributed electrons and ions. At equilibrium, the quasi-neutrality condition is $Z_1 n_{10}+n_{e0}=Z_2 n_{20}+n_{i0}$,
where $n_{10}$, $n_{e0}$, $n_{20}$, and $n_{i0}$ are the equilibrium number densities of warm negatively
charged dust, non-extensive $q$-distributed electrons, positively charged dust particles, and isothermal ions,
respectively. The normalized governing equations of the DAWs can be represented by
\begin{eqnarray}
&&\frac{\partial n_1}{\partial t}+\frac{\partial}{\partial x}(n_1 u_1)=0,
\label{eq1}\\
&&\frac{\partial u_1}{\partial t} + u_1\frac{\partial u_1 }{\partial x} + 3\sigma_1  n_1\frac{\partial n_1 }{\partial x}=\frac{\partial \phi}{\partial x},
\label{eq2}\\
&&\frac{\partial n_2}{\partial t}+\frac{\partial}{\partial x}(n_2 u_2)=0,
\label{eq3}\\
&&\frac{\partial u_2}{\partial t} + u_2\frac{\partial u_2 }{\partial x}+ 3\sigma_2  n_2\frac{\partial n_2 }{\partial x}=-\alpha \frac{\partial \phi}{\partial x},
\label{eq4}\\
&&\frac{\partial^2 \phi}{\partial x^2}=(\mu_i+\beta-1)n_e-\mu_i n_i+n_1-\beta n_2.
\label{eq5}
\end{eqnarray}
In the above equations, $n_1$ ($n_2$) is the number density of negatively (positively) charged dust particles
normalized by its equilibrium value $n_{10}$ ($n_{20}$); $u_1(u_2)$ is the negative (positive) charged dust fluid speed
normalized by $C_1=(Z_1 T_i/m_1)^{1/2}$, and the electrostatic wave potential $\phi$
is normalized by $T_i/e$ (where $e$ is the magnitude of an electron charge);
$T_1$, $T_2$, $T_i$, and $T_e$ are the temperatures of  negatively charged dust,
positively charged dust, non-extensive $q$-distributed ions, and electrons, respectively; the time and space variables are normalized by
${\omega^{-1}_{pd1}}=(m_1/4\pi Z^2_1e^2 n_{10})^{1/2}$ and $\lambda_{Dd1}=(T_{i}/4 \pi Z_1 e^2 n_{10})^{1/2}$, respectively; where
$\sigma_1=T_1/Z_1 T_i$, $\sigma_2=T_2 m_1/Z_1T_i m_2$, $\alpha=Z\mu$, $Z=Z_2/Z_1$, $\mu=m_1/m_2$,
$\mu_i=n_{i0}/Z_1n_{10}$, $\beta=ZR$, $R=n_{20}/n_{10}$, and $\delta=T_i/T_e$.
The expression for the number density of  non-extensive $q$-distributed electrons and ions following the non-extensive $q$-distribution can be expressed as
 \begin{eqnarray}
&&n_e= \left[1+(q-1)\delta \phi\right]^\frac{q+1}{2(q-1)},
\label{eq6}\\
&&n_i= \left[1-(q-1)\phi \right]^\frac{q+1}{2(q-1)}.
\label{eq7}
\end{eqnarray}
Now, by substituting Eqs. (\ref{eq6}) and (\ref{eq7}) into Eq. (\ref{eq5}), and expanding it up to third order, we get
\begin{eqnarray}
&&\hspace*{-1cm}\frac{\partial^2 \phi}{\partial x^2}=\beta-1+n_1-\beta n_2+\gamma_1 \phi+\gamma_2\phi^2+\gamma_3 \phi^3\cdot\cdot\cdot\cdot,
\label{eq8}
\end{eqnarray}
where
\begin{eqnarray}
&&\gamma_1=\frac{(q+1)[\delta(\mu_i+\beta-1)+\mu_i]}{2},
\nonumber\\
&&\gamma_2=\frac{(q+1)(3-q)[{\delta^2(\mu_i+\beta-1)-\mu_i}]}{8},
\nonumber\\
&&\gamma_3=\frac{(q+1)(q-3)(3q-5)[{\delta^3(\mu_i+\beta-1)+\mu_i}]}{48}.
\nonumber\
\end{eqnarray}
To study the modulation of the DAWs plasma system, we will derive the NLS equation by employing the reductive perturbation
method. So, we first introduce the  stretched co-ordinates \cite{Chowdhury2018}
\begin{eqnarray}
&&\xi={\epsilon}(x-v_gt),
\label{eq9}\\
&&\tau={\epsilon}^2 t,
\label{eq10}
\end{eqnarray}
where $v_g$ is the envelope group velocity to be determined later, and $\epsilon$ is a
smallness parameter ($0<\epsilon<1$). Then, we can write a general expression for the dependent variables as
\begin{eqnarray}
&&G(x,t)=G_0 +\sum_{m=1}^{\infty}\epsilon^{(m)}\sum_{l=-\infty}^{\infty}G_{l}^{(m)}(\xi,\tau) ~\mbox{exp}(i l\Theta),
\label{eq11}
\end{eqnarray}
where $G_l^{(m)}=[n_{1l}^{(m)}, u_{1l}^{(m)}, n_{2l}^{(m)}, u_{2l}^{(m)}, \phi_l^{(m)}]^T$,
$G_0=[1, 0, 1, 0, 0]^T$, $\Theta=(k x -\omega t)$, and $k$ ($\omega$) is the fundamental carrier wave number (frequency).
All elements of $G_l^{(m)}$ satisfy the reality condition $G_{-l}^{(m)}=G_l^{*(m)}$, where the asterisk indicates the complex conjugate.
The derivative operators in the above equations are treated as follows:
\begin{eqnarray}
&&\frac{\partial}{\partial t}\rightarrow\frac{\partial}{\partial t}-\epsilon v_g \frac{\partial}{\partial\xi}+\epsilon^2\frac{\partial}{\partial\tau},
\label{eq12}\\
&&\frac{\partial}{\partial x}\rightarrow\frac{\partial}{\partial x}+\epsilon\frac{\partial}{\partial\xi}.
\label{eq13}
\end{eqnarray}
Substituting Eqs. (\ref{eq11}) and (\ref{eq13}) into equations Eqs. (\ref{eq1})$-$(\ref{eq4}), and (\ref{eq8}) and
equating the coefficients of different powers of $\epsilon$ for $m=l=1$, one obtains
\begin{eqnarray}
&&n_{11}^{(1)}=\frac{k^2}{S}\phi_1^{(1)},~~~~~ u_{11}^{(1)}=\frac{k\omega}{S}\phi_1^{(1)},\nonumber\\
&&n_{21}^{(1)}=\frac{\alpha k^2}{A}\phi_1^{(1)},~~~~~u_{21}^{(1)}=\frac{\alpha k\omega}{A}\phi_1^{(1)},
\label{eq14}
\end{eqnarray}
where $S=\lambda k^2-\omega^2$, $A=\omega^2-\theta k^2$, $\lambda=3 \sigma_1$ and $\theta=3 \sigma_2$. We thus obtain the dispersion relation for DAWs
\begin{eqnarray}
&&\omega^2=\frac{k^2M\pm k^2 \sqrt{M^2-4 GH}}{2G},
\label{eq15}
\end{eqnarray}
where $M=(\theta k^2+\lambda k^2+\theta \gamma_1+\lambda\gamma_1+\alpha\beta+1), G=(k^2+\gamma_1 )$, and
$H=(\theta\lambda k^2+\theta\gamma_1\lambda+\theta+\alpha \beta\lambda).$
In Eq. (\ref{eq15}), the positive (negative) sign corresponds to the fast (slow) modes when $M^2>4GH$.
The variations of fast mode ($\omega_f$) vs carrier wave number ($k$) for different values are depicted in Fig. \ref{1Fig:F1}. The value of
$\omega_f$ increases rapidly with $k$ for $k<1.0$, and becomes nearly steady for $k>1.0$. Also, with increase of the
value of ratio of the charge state ($Z$), the value of $\omega_f$ increases. So, $\omega_f$ is increased (decreased)
with the increase of positively (negatively) charged dust charge. The variations of slow mode ($\omega_s$) vs $k$ for different
values of $Z$ is depicted in Fig. \ref{1Fig:F2}. The value of $\omega_s$ increases linearly with $k$ passing through the
origin, and with the increase of the value of $Z$, the value of $\omega_s$ decreases, which is totally opposite with
the behaviour of $\omega_f$. So, $\omega_s$ is increased (decreased) with the increase of negatively (positively) charged dust charge.
Basically, both dust species oscillate in phase with electrons and ions in case of the fast DAWs.
But only one dust species oscillate in phase with electrons and ions in case of the slow DAWs.
The second-order ($m=2$) reduced equations with ($l=1$) give the group velocity $v_g$ (along with the compatibility condition)
\begin{eqnarray}
&&v_g=\frac{\partial \omega}{\partial k}=N.
\label{eq16}
\end{eqnarray}
The amplitude of the second-order harmonics are found to be proportional to $|\phi_1^{(1)}|^2$
\begin{eqnarray}
&&n_{12}^{(2)}=C_1|\phi_1^{(1)}|^2,~~~n_{10}^{(2)}=C_6 |\phi_1^{(1)}|^2,
\nonumber\\
&&u_{12}^{(2)}=C_2 |\phi_1^{(1)}|^2,~~~~u_{10}^{(2)}=C_7|\phi_1^{(1)}|^2,
\nonumber\\
&&n_{22}^{(2)}=C_3|\phi_1^{(1)}|^2,~~~~n_{20}^{(2)}=C_8 |\phi_1^{(1)}|^2,
\nonumber\\
&&u_{22}^{(2)}=C_4 |\phi_1^{(1)}|^2,~~~~u_{20}^{(2)}=C_9|\phi_1^{(1)}|^2,
\nonumber\\
&&\phi_2^{(2)}=C_5 |\phi_1^{(1)}|^2,~~~~\phi_0^{(2)}=C_{10} |\phi_1^{(1)}|^2.
\label{eq17}
\end{eqnarray}
Finally, the third harmonic modes ($m=3$) and ($l=1$),  with the help of  Eqs. (\ref{eq14})$-$(\ref{eq17}),
give a system of equations, which can be reduced to the following  NLS equation
\begin{eqnarray}
&&i\frac{\partial \Phi}{\partial \tau}+P\frac{\partial^2 \Phi}{\partial \xi^2}+Q|\Phi|^2\Phi=0,
\label{eq18}
\end{eqnarray}
where $\Phi=\phi_1^{(1)}$ for simplicity. The dispersion coefficient $P$
and the nonlinear coefficient $Q$ are given in the appendix.
\section{Stability and Rogue waves}
\label{sec:Stability}
\begin{figure}[t!]
\centering
\includegraphics[width=70mm]{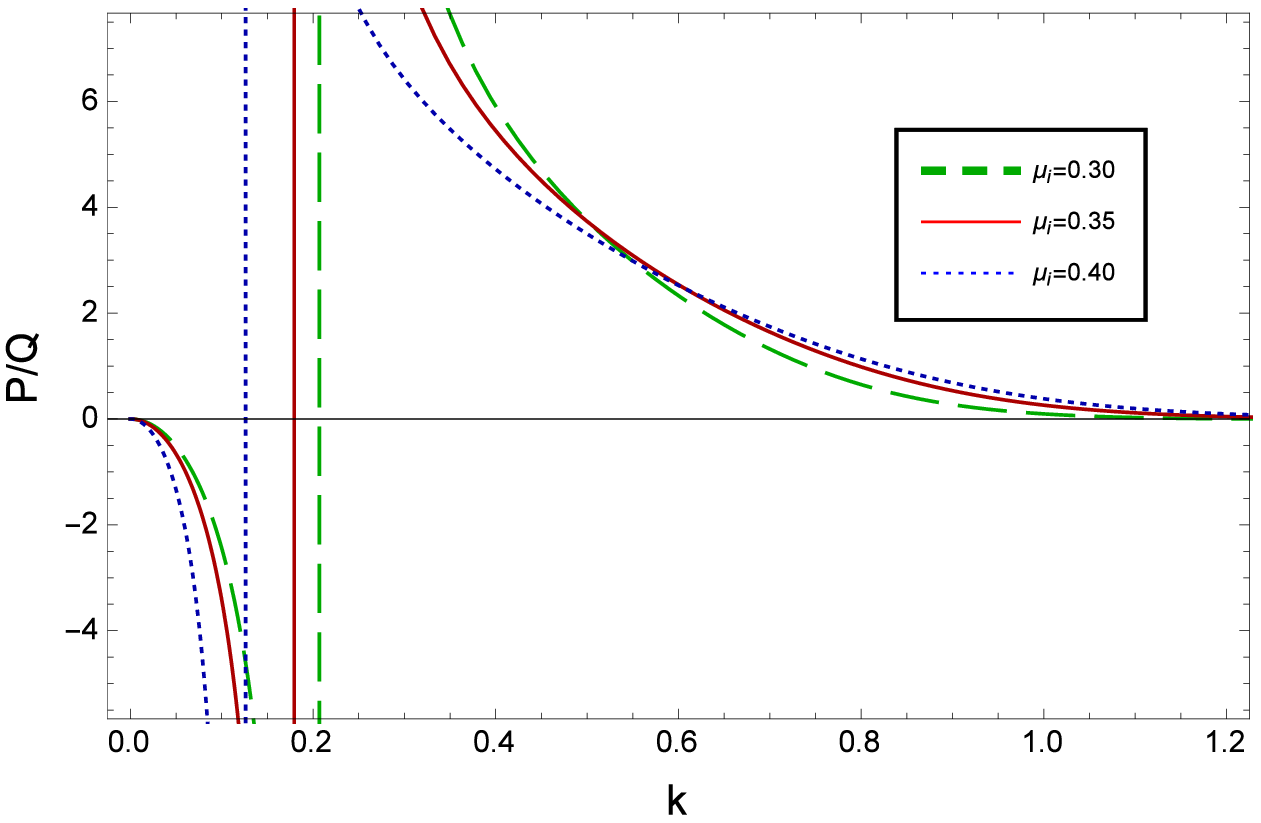}
\caption{The variation of $P/Q$ with $k$ for different values of $\mu_i$;
  along with $Z=0.01$, $\mu=150$, $R=0.1$, $\delta=0.3$,  $q=1.5$, $\sigma_1=0.0001$, $\sigma_2=0.001$, and $\omega_f$.}
\label{1Fig:F3}
\vspace{0.8cm}
\includegraphics[width=70mm]{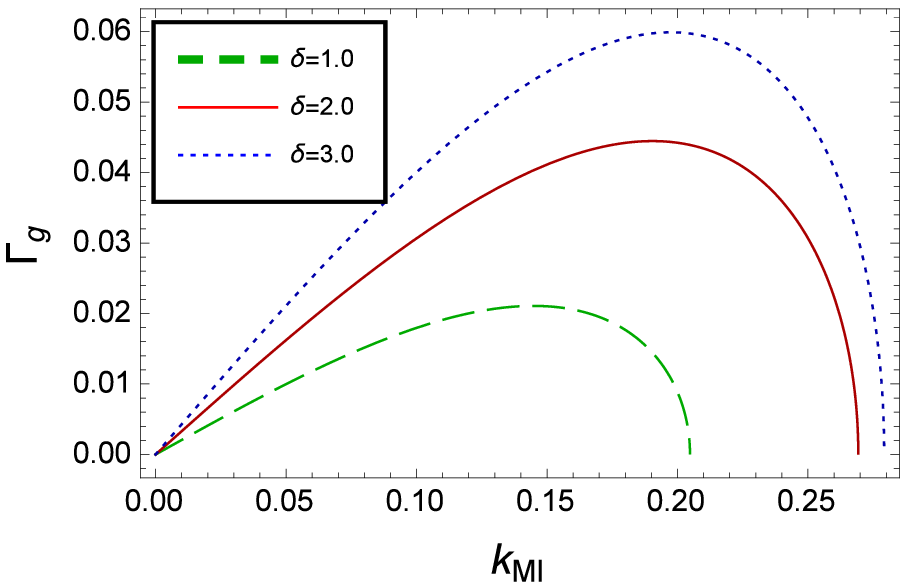}
\caption{The variation of MI growth rate $(\Gamma_g)$ with $K_{MI}$ for $\delta$;
  along with $Z=0.01$, $\mu=150$, $R=0.1$, $\mu_i=0.4$,  $q=1.5$, $\sigma_1=0.0001$,
  $\sigma_2=0.001$, $k=0.3$, $\Phi_0=0.5$, and $\omega_f$.}
 \label{1Fig:F4}
\end{figure}

The nonlinear evolution of the DAWs typically depends on the coefficients of dispersion
term $P$ and nonlinear term $Q$, which are function of the various plasma parameters such
as $Z$, $\mu$, $R$, $\sigma_1$, $\sigma_2$, $\mu_i$, $\delta$, and $q$. Thus, these plasma parameters significantly
controlled the stability conditions of the DAWs. If $P/Q<0$, DAWs are modulationally stable
and for the case $P/Q>0$, DAWs are modulationally unstable against external perturbations
\cite{Sukla2002,Fedele2002} and simultaneously
when $P/Q>0$ and ${k_{MI}}<k_c$, the growth rate ($\Gamma_g$) of  MI is
given \cite{Shalini2015}  by
\begin{eqnarray}
&&\Gamma_g=|P|~{k^2_{MI}}\sqrt{\frac{k^2_{c}}{k^2_{MI}}-1},
\label{eq19}
\end{eqnarray}
where ${k_{MI}}$ is the perturbation wave number and the critical value of the wave number of modulation
$k_c=\sqrt{2Q{|\Phi_o|}^2/P}$, and $\Phi_o$ is the amplitude of the carrier waves.
Hence, the maximum value $\Gamma_{g(max)}$ of $\Gamma_g$ is obtained at ${k_{MI}}=k_c/\sqrt{2}$,
and is given by $\Gamma_{g(max)}=|Q||\Phi_0|^2$. The stability of the profile is investigated by
depicting the ratio of $P/Q$ (for $\omega_f$) against $k$ for different values of ion number density
($\mu_i$) with fixed values of other physical parameters in Fig. \ref{1Fig:F3}. If $P/Q>0$, the DAWs
are modulationally unstable, and for the case $P/Q<0$, the DAWs are modulationally stable against
external perturbations. When $P/Q\rightarrow\infty$, the corresponding value of $k$ ($=k_c$) is
called critical or threshold wave number for the onset of MI. With the increase of the value
of $\mu_i$, the value of $k_c$ decreases. The threshold $k_c$ differentiates the stable and
unstable region, which decreases with $\mu_i$. The growth rate ($\Gamma_g$) increases with the
increase of ${k_{MI}}$. For a particular value of ${k_{MI}}$, the growth rate ($\Gamma_g$)
is obtained at its critical value ($\Gamma_g\equiv \Gamma_{gc}$). Hence, with the increase of the
value of ${k_{MI}}$, the growth rate ($\Gamma_g$) decreases significantly. Moreover,
with the increase of the value of $\delta$, the critical growth rate($\Gamma_{gc}$) also increases.
\begin{figure}[t!]
\centering
\includegraphics[width=70mm]{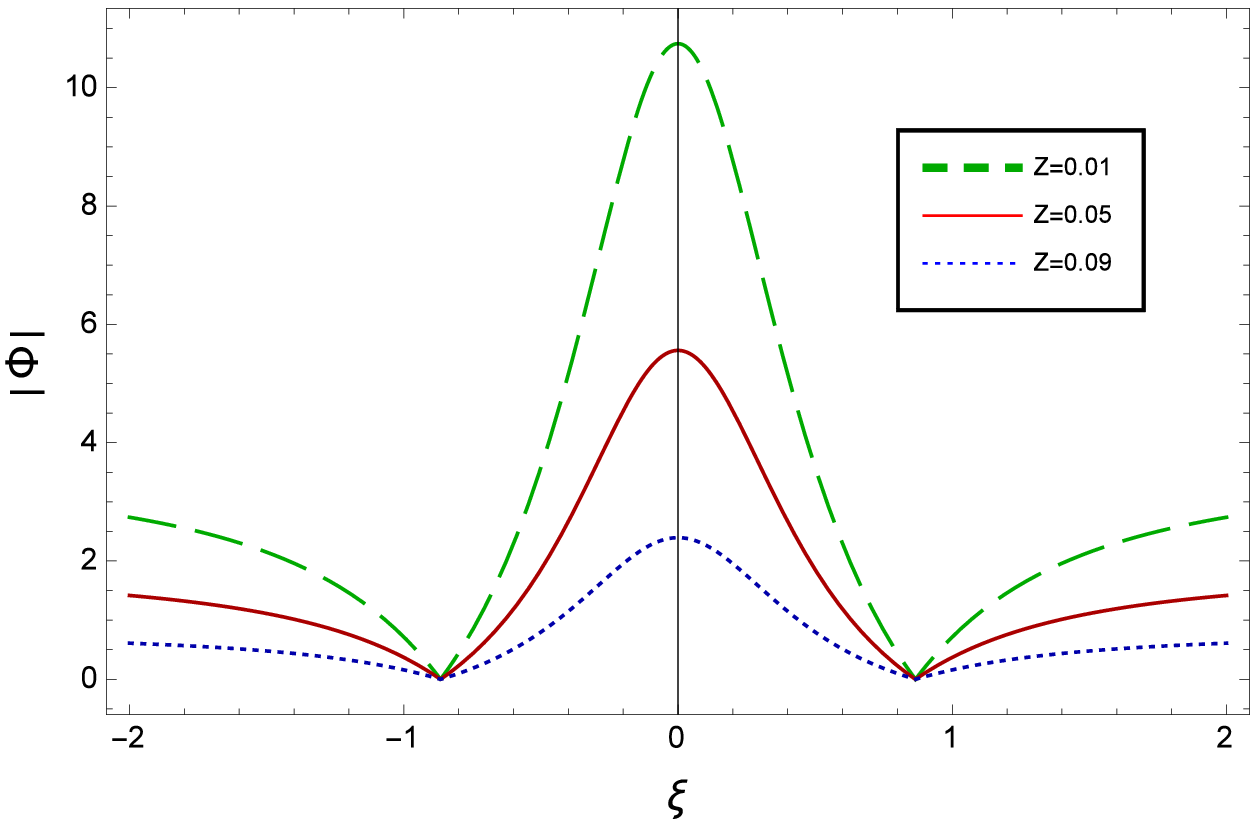}
\caption{The variation of $|\Phi|$ with $\xi$ for different values of $Z$;
  along with $\mu=150$, $R=0.1$, $\mu_i=0.4$,  $q=1.5$, $\sigma_1=0.0001$,
  $\sigma_2=0.001$, $k=0.3$, $\tau=0$, and $\omega_f$.}
\label{1Fig:F5}
\vspace{0.8cm}
\includegraphics[width=70mm]{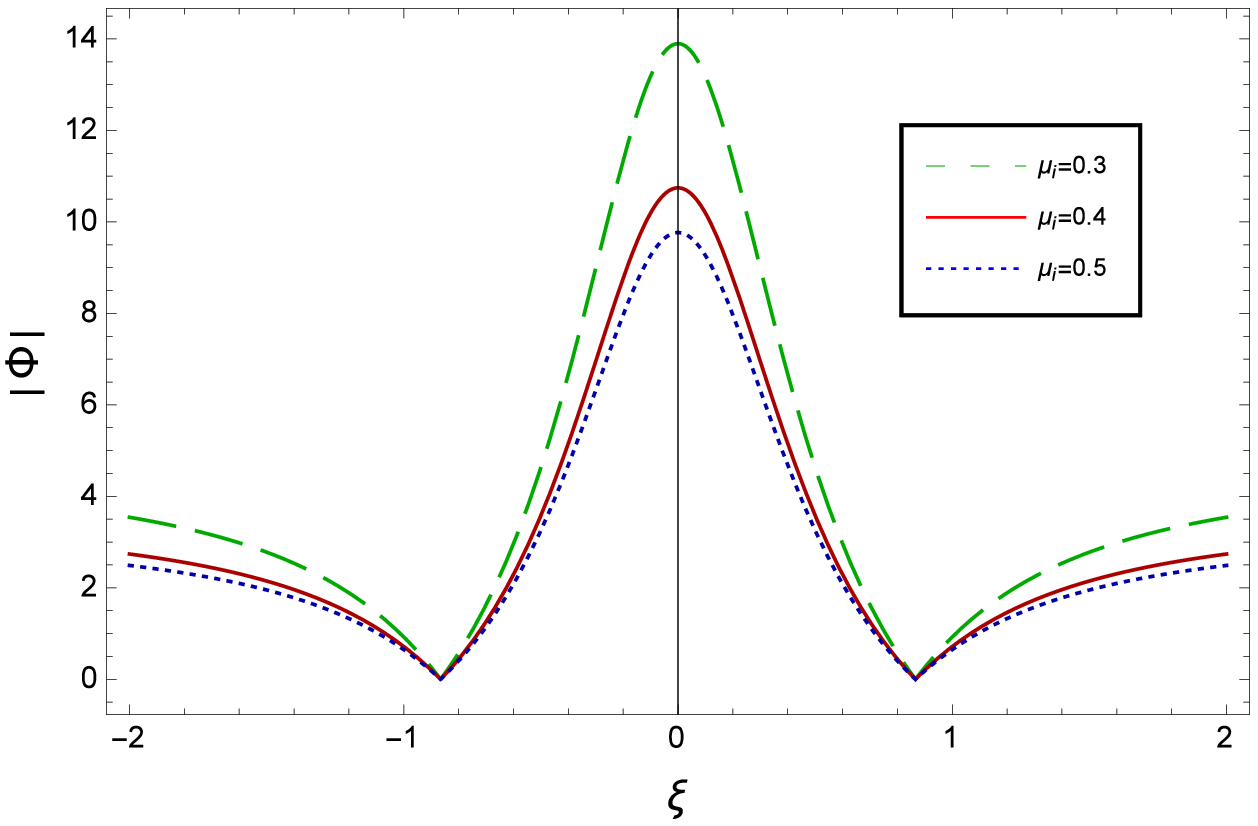}
\caption{The variation of $|\Phi|$ with $\xi$ for different values of $\mu_i$;
  along with $Z=0.01$, $\mu=150$, $R=0.1$,   $q=1.5$, $\sigma_1=0.0001$,
  $\sigma_2=0.001$, $k=0.3$, $\tau=0$, and $\omega_f$.}
 \label{1Fig:F6}
\end{figure}
\begin{figure}[t!]
\centering
\includegraphics[width=70mm]{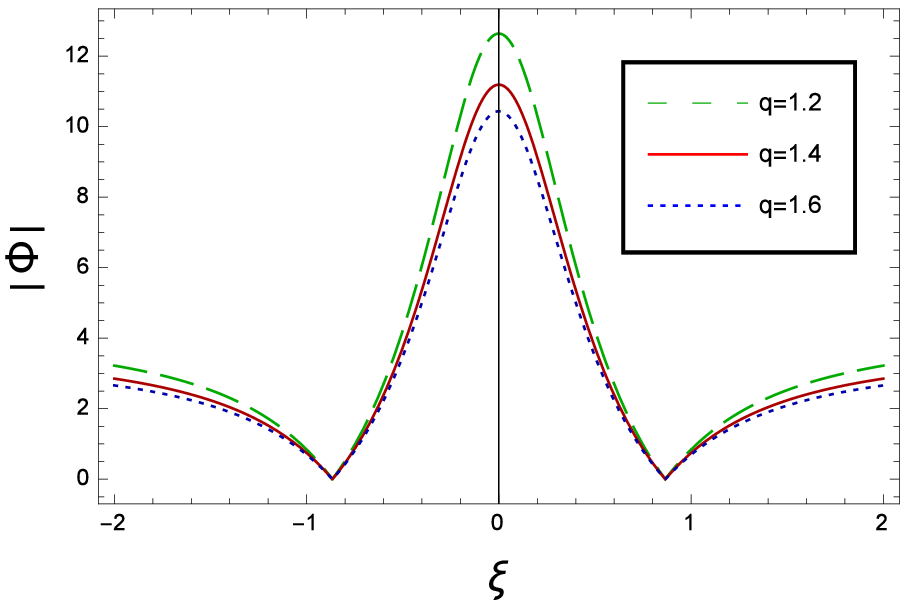}
\caption{The variation of $|\Phi|$ with $\xi$ for different values of $q$;
  along with $Z=0.01$, $\mu=150$, $R=0.1$,   $\mu_i=0.4$, $\sigma_1=0.0001$,
  $\sigma_2=0.001$, $k=0.3$, $\tau=0$, and $\omega_f$.}
\label{1Fig:F7}
\vspace{0.8cm}
\includegraphics[width=70mm]{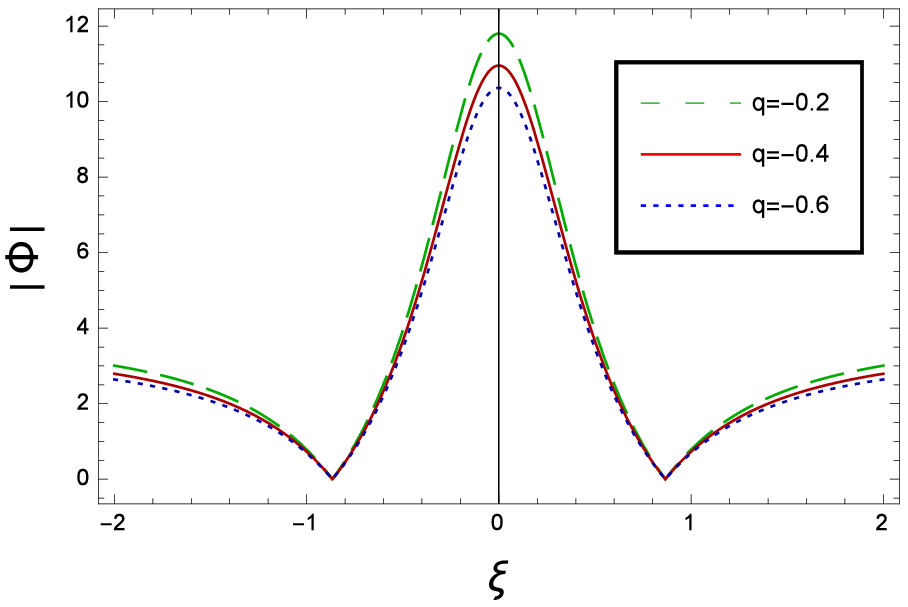}
\caption{The variation of $|\Phi|$ with $\xi$ for different values of $q$;
  along with $Z=0.01$, $\mu=150$, $R=0.1$,   $\mu_i=0.4$, $\sigma_1=0.0001$,
  $\sigma_2=0.001$, $k=0.3$, $\tau=0$, and $\omega_f$.}
 \label{1Fig:F8}
\end{figure}
The  rogue wave (rational solution) of the NLS equation (\ref{eq18}) in the unstable
region ($P/Q>0$) can be written \cite{Ankiewiez2009} as
\begin{eqnarray}
&&\Phi(\xi,\tau)=\sqrt{\frac{2P}{Q}} \left[\frac{4(1+4iP\tau)}{1+16P^2\tau^2+4\xi^2}-1 \right]\mbox{exp}(i2P\tau).
\label{eq20}
\end{eqnarray}
The solution (\ref{eq20}) anticipates the concentration of the DAW rogue waves (DARWs) energy into a small
region that is caused by the nonlinear behavior of the plasma medium.
The variation of $|\Phi|$ against $\xi$ is illustrated in Fig. \ref{1Fig:F5}$-$\ref{1Fig:F8}.
Fig. \ref{1Fig:F5} expresses the dependency of the
electrostatic potential against ratio of the both dust charge state ($Z$). From this view point,
it can be deduced that more massive negatively charged dust particle enhanced the nonlinearity
of the system, which leads to generate energetic (by increasing hight and width of the rogue
structures) profile. The amplitude and width of the rogue structures decrease in Fig. \ref{1Fig:F6},
which also illustrates the variation of electrostatic potential against ion number density ($\mu_i$).

Rogue structures are so much sensitive to the change of the non-extensive parameter ($q$). When $q$ is
positive, with the increase of the value of $q$, the shape of the rogue structures are changed by
decreasing the amplitude and width [please see Fig. \ref{1Fig:F7}]. But when $q<0$, the amplitude and width of the
rogue structures decrease with the decrease of the value of $q$ [please see Fig. \ref{1Fig:F8}].
So, the rogue structures are crucially depend on the sign of $q$.
\section{Discussion}
\label{sec:Discussion}
We have investigated the amplitude modulation of DAWs in an unmagnetized four-component plasma system consisting
of inertial warm negatively charged dust particles and positively charged dust particles as well as non-extensive
$q$-distributed electrons and ions.
The NLS equation, which governs the evolution of nonlinear DAWs, is derived  by employing the reductive perturbation method.
The results, which have been found from this theoretical investigation, can be summarized as follows:
\begin{enumerate}
\item{The fast mode increases rapidly with $k$ for $k<1.0$, and becomes nearly steady for $k>1.0$. Also, with increase of $Z$, the fast mode increases.}
\item{The slow mode increases linearly with $k$ passing through the origin, and with the increase of $Z$, the slow mode decreases, which is totally opposite with the behaviour of the fast mode.}
\item{The DAWs will be modulationally unstable (stable) with $k$ in which the ratio $P/Q$ is positive (negative).}
\item{The growth rate ($\Gamma_g$) increases with the increase of ${k_{MI}}$. After obtaining the critical growth rate ($\Gamma_{gc}$), the growth rate ($\Gamma_g$) decreases significantly with ${k_{MI}}$.
   Moreover, $\Gamma_{gc}$ also increases with $\delta$.}
\item{The amplitude and width of the DAWs decrease with the increase of both $Z$ and $\mu_i$.}
\item{The amplitude and width of the rogue waves decrease (increase) as we
 increase the value of $q>0$ ($q<0$).}
\end{enumerate}
We may hope that the results of this theoretical investigation could be helpful for understanding the
 nonlinear electrostatic structures in astrophysical (viz. supernova, planetary rings, earth's ionosphere \cite{Chowdhury2017}, etc.) and laboratory environments (laser plasma \cite{Chowdhury2018}).
\newpage
\section{Appendix: Expressions of the  Coefficients}
\label{sec:Appendix}
\begin{eqnarray}
&&\hspace{-3.0cm}N= \frac{k^2(\lambda A^2 + \alpha \beta \theta S^2)+ \omega^2 (A^2+\alpha \beta S^2)-2 S^2 A^2-SA(A-\alpha \beta S )}{2k \omega( A^2 + \alpha \beta S^2 )},\nonumber\\
&&\hspace{-3.0cm}C_1=\frac{2C_5 k^2 S^2  -(3 \omega^2 k^4+\lambda k^6)}{2S^3},~~~~~C_2=\frac{\omega C_1 S^2 -\omega k^4}{k S^2}, \nonumber\\
&&\hspace{-3.0cm}C_3=\frac{3\alpha^2\omega^2 k^4 +\theta \alpha^2 k^6+2\alpha C_5 A^2 k^2}{2A^3},~~~~~C_4=\frac{ \omega C_3 A^2-\omega \alpha^2 k^4 }{k A^2},  \nonumber\\
&&\hspace{-3.0cm}C_5=\frac{A^3 (3\omega^2k^4+\lambda k^6)-2\gamma_2 A^3 S^3+\beta S^3(3\alpha^2 \omega^2 k^4+\theta\alpha^2 k^6)}{2S^2k^2 A^3+2A^3 S^3 (4k^2+\gamma_1)-2\alpha\beta A^2 k^2S^3}, \nonumber\\
&&\hspace{-3.0cm}C_6=\frac{2 v_g \omega k^3+\lambda k^4+k^2\omega^2-C_{10}S^2}{S^2(v^2_g-\lambda)},~~~~~C_7=\frac{v_g C_6 S^2-2\omega k^3}{S^2}, \nonumber\\
&&\hspace{-3.0cm}C_8=\frac{2 v_g \omega\alpha^2 k^3+\theta \alpha^2k^4+\alpha^2 k^2 \omega ^2+\alpha C_{10}A^2}{A^2(v^2_g - \theta )},~~~~~C_9=\frac{v_g C_8 A^2 -2\omega\alpha^2 k^3}{A^2}, \nonumber\\
&&\hspace{-3.0cm}C_{10} =\frac{A^2( 2 v_g \omega k^3+\lambda k^4+k^2 \omega ^2 )(  v^2_g - \theta )+2\gamma_2 A^2 S^2(v^2_g -\theta )( v^2_g -\lambda)-\beta S^2( 2 v_g \omega \alpha^2 k^3+\theta \alpha^2 k^4+\alpha^2 k^2 \omega ^2)(v_g^2 -\lambda)}{\alpha \beta A^2 S^2 (v^2_g -\lambda)+A^2 S^2 (v^2_g - \theta)- \gamma_1 A^2 S^2(v^2_g - \theta)(v^2_g -\lambda)}, \nonumber\\
&&\hspace{-3.0cm}P =\frac{1}{2 AS\omega k^2(A^2+\alpha \beta S^2)}[A^3((v_g \omega -\lambda k)(\lambda k^3-2\omega v_g k^2+k \omega^2-kS)+(v_g k-\omega)
                                              (\lambda \omega k^2 - 2v_g k \omega^2+\omega^3-k v_g S)).\nonumber\\
&&\left.-\alpha \beta S^3((v_g \omega -\theta k)(\theta k^3-2\omega v_g k^2+k\omega^2+kA)+(v_g k-\omega)(\theta\omega k^2 - 2v_g k\omega^2+\omega^3+kv_gA))-A^3S^3\right],\nonumber\\
&&\hspace{-3.0cm}Q=\frac{A^2S^2}{2\omega k^2(A^2+\alpha \beta S^2)}\left[2\gamma_2(C_5+C_{10})+3\gamma_3-\frac{2\omega k^3(C_2+C_{7})}{S^2}-\frac{2\alpha \beta \omega k^3(C_4+C_9)}{A^2} \right.\nonumber\\
&&\left.~~~~~~~~~~~~~~~~~~-\frac{(\omega^2k^2+\lambda k^4)(C_1+C_6)}{S^2} -\frac{(\alpha \beta k^2\omega^2+\alpha \beta \theta k^4)(C_3+C_8)}{A^2}\right].\nonumber\
\end{eqnarray}


\section*{References}
\begin{thebibliography}{100}

\bibitem{Shukla2002} P. K. Shukla and A. A. Mamun, Introduction to Dusty Plasma Physics (Institute of Physics Publishing Ltd., Bristol, 2002).

\bibitem{Mamun20002} A. A. Mamun and P. K. Shukla, Geophys. Res. Lett. {\bf29}, 1870 (2002).

\bibitem{Chowdhury2017a} N. A. Chowdhury, A. Mannan, and A. A. Mamun, Phys. plasmas {\bf24} 113701 (2017).

\bibitem{El-Taibany2013} W. F. El-Taibany, Phys. Plasmas {\bf20}, 093701 (2013).

\bibitem{Zhukhovitskii2015} D. I. Zhukhovitskii, Phys. Rev. E {\bf92}, 023108 (2015).

\bibitem{El-Taibany2008}W. F. El-Taibany, I. Kourakis, and Miki Wadati, Plasma Phys. Control. Fusion {\bf50}, 074003 (2008).

\bibitem{El-Labany2013} S. K. El-Labany, W. F. El-Taibany, and E. E. Behery, Phys. Rev. E {\bf88}, 023108 (2013).

\bibitem{Ali1998} F. S. Ali, M. A. Ali, R. A. Ali, and I. I. Lnculet, J. Electrostatics {\bf45}, 139 (1998).

\bibitem{Zhao2002}  H. Zhao, G. S. P. Castle, and I. I. Lnculet, J. Electrostatics {\bf55}, 261 (2002).

\bibitem{Mamun2015} A. A. Mamun, M. Ferdousi, and S. Sultana, Phys. Scr. {\bf90}, 088011 (2015).

\bibitem{Mamun2008} A. A. Mamun, Phys. Rev. E {\bf77}, 026406 (2008).

\bibitem{Mamun2016} A. A. Mamun and R. Schlickeiser, Phys. Plasmas {\bf23}, 034502 (2016).

\bibitem{Mendis1991} D. A. Mendis and M. Horanyi, Geophys. Mongr. Ser. {\bf61}, 17 (1991).

\bibitem{Chow1993}V. W. Chow, D. A. Mendis, and M. Rosenberg, J. Geophys. Res. {\bf98}, 19065 (1993).

\bibitem{Horanyi1993} M. Horanyi, G. E. Morfill, and E. Grun, Nature, {\bf363}, 144 (1993).

\bibitem{Shukla2006} P. K. Shukla and M. Rosenberg, Phys. Scr. {\bf73}, 196 (2006).

\bibitem{Angelo2004} N. D' Angelo, J. Phys. D: Appl. Phys. {\bf37}, 860 (2004).

\bibitem{Mamun2002} A. A. Mamun and P. K. Shukla, Geophys. Res. Lett. {\bf29}, 1870 (2002).

\bibitem{Ismael2017} I. Driouch and H. Chatei, Eur. Phys. J. D {\bf71}, 9 (2017).

\bibitem{Rafat2015} A. Rafat, M.M. Rahman, M.S. Alam, A.A. Mamun, Astrophys. Space Sci. {\bf358}, 19 (2011).

\bibitem{Shalini2015} Shalini, N. S. Saini, and A. P. Misra, Phys. Plasmas {\bf22}, 092124 (2015).

\bibitem{Renyi1955} A. Renyi, Acta Math. Acad. Sci. Hung. {\bf6}, 285 (1955).

\bibitem{Tsallis1988} C. Tsallis, J. Stat. Phys. {\bf52}, 479 (1988).

\bibitem{Plastino1993} A. R. Plastino and A. Plastino, Phys. Lett. A {\bf174}, 384 (1993).

\bibitem{Gervino2012} G. Gervino, A. Lavagno, and D. Pigato, Cent. Eur. J. Phys. {\bf10}, 594 (2012).

\bibitem{Feron2008} C. Feron and J. Hjorth, Phys. Rev. E {\bf77}, 022106 (2008).

\bibitem{Asbridge1968} J. R. Asbridge, S. J. Bame, and I. B. Strong, J. Geophys. Res. {\bf73}, 5777, (1968).

\bibitem{Krimigis1983} S. M. Krimigis, J. F. Carbary, E. P. Keath, T. P. Armstrong, L. J. Lanzerotti, and G. Gloeckler, J. Geophys. Res. {\bf88}, 8871 (1983).

\bibitem{Vladimirov2004} S. V. Vladimirov and K. J. Ostrikov, Phys. Rep. {\bf393}, 175 (2004).

\bibitem{Akhmediev2009} N. Akhmediev, A. Ankiewicz, and J. M. Soto-Crespo, Phys. Rev. E {\bf80}, 026601 (2009).

\bibitem{Moslem2011} W. M. Moslem, R. Sabry, S. K. El-Labany, and P. K. Shukla, Phys. Rev. E {\bf84}, 066402 (2011).

\bibitem{MSE2011} W. M. Moslem, P. K. Shukla, and B. Eliasson, Europhys. lett. {\bf96}, 25002 (2011).

\bibitem{Solli2007} D. R. Solli, C. Ropers, P. Koonath, and B. Jalali, Nature (London) {\bf450}, 1054 (2007).

\bibitem{Kibler2010} B. Kibler, J. Fatome, C. Finot {\it et al.}, Nat. Phys. (London) {\bf6}, 790 (2010).

\bibitem{Bludov2009} Yu. V. Bludov, V. V. Konotop, and N. Akhmediev, Phys. Rev. A {\bf80}, 033610 (2009).

\bibitem{Stenflo2010} L. Stenflo and M. Marklund, J. Plasma Phys., {\bf76}, 293 (2010).

\bibitem{Chowdhury2017} N.A. Chowdhury, A. Mannan, M.M. Hasan, A.A. Mamun,  Chaos  27 093105 (2017).

\bibitem{Turing1952} A. M. Turing, Philos. Trans. R. Soc. London B {\bf237}, 37 (1952).

\bibitem{Yan2010} Z. Yan, Commun. Theor. Phys. {\bf54}, 947 (2010).

\bibitem{Bains2013} A.S. Bains, M. Tribeche, C.S. Ng, Astrophys. Space Sci. {\bf343} 621 (2013).

\bibitem{Chowdhury2018} N. A. Chowdhury, M. M. Hasan, A. Mannan, and A. A. Mamun, Vacuum {\bf147} 31 (2018).

\bibitem{Sukla2002} R. Fedele, H. Schamel, and P.K. Sukla, Phys. Scr.  {\bf98}, 18 (2002).

\bibitem{Fedele2002} R. Fedele, Phys. Scr. {\bf65}, 502 (2002).

\bibitem{Ankiewiez2009} A. Anikiewicz, N. Devine, and N. Akhmediev, Phys. Lett. A {\bf373}, 3997 (2009).

\end{thebibliography}
\end{document}